\documentclass{iaus}
\usepackage{graphicx}
\usepackage{array}
\usepackage{amssymb}
\usepackage{color}
\usepackage{bm}

\title[Reference Frames and the Physical Gravito-Electromagnetic Analogy] 
{Reference Frames and the Physical Gravito-Electromagnetic Analogy}

\author[L. Filipe O. Costa \& Carlos A. R. Herdeiro]   
{L. Filipe O.  Costa$^1$
 \and Carlos A. R. Herdeiro$^2$}

\affiliation{$^{1,2}$Centro de F{\'\i}sica do Porto e Departamento de F{\'\i}sica da Universidade do Porto \\ Rua do Campo Alegre 687, 4169-007 Porto, Portugal \\$^1$email: {\tt filipezola@fc.up.pt}, $^2$email: {\tt crherdei@fc.up.pt}\\Illustrations by Rui Quaresma (\tt quaresma.rui@gmail.com)}

\pubyear{2009}
\volume{261}  
\pagerange{xxx-xxx}
\jname{Relativity in Fundamental Astronomy}
\editors{S. A. Klioner, P. K. Seidelman \& M. H. Soffel, eds.}
\begin{document}

\maketitle

\begin{abstract}
The similarities between linearized gravity and electromagnetism are
known since the early days of General Relativity. Using an exact approach
based on tidal tensors, we show that such analogy holds only on very
special conditions and depends crucially on the reference frame. This
places restrictions on the validity of the {}``gravito-electromagnetic''
equations commonly found in the literature.


\end{abstract}



\section{Gravito-electromagnetic analogy based on tidal tensors}

The topic of the gravito-electromagnetic analogies has a long story,
with different analogies being unveiled throughout the years. Some
are purely formal analogies, like the splitting of the Weyl tensor
in electric and magnetic parts, e.g. \cite{Maartens}; but others
(e.g \cite{DSX,CostaHerdeiro2008,Jantzen,Natario,Ruggiero:2002})
stem from certain physical similarities between the gravitational
and electromagnetic interactions. The linearized Einstein equations
(see e.g. \linebreak  \cite{DSX,Ruggiero:2002,Gravitation and Inertia}), in the
harmonic gauge $\bar{h}_{\alpha\beta}^{\,\,\,\,\,\,,\beta}=0$, take
the form $\square\bar{h}^{\alpha\beta}=-16\pi T^{\alpha\beta}/c^{4}$,
similar to Maxwell equations in the Lorentz gauge: $\square A^{\beta}=-4\pi j^{\beta}/c$.
That suggests an analogy between the trace reversed time components
of the metric tensor $\bar{h}_{0\alpha}$ and the electromagnetic
4-potential $A_{\alpha}$. Defining the 3-vectors usually dubbed gravito-electromagnetic
fields, the time components of these equations may be cast in a Maxwell-like
form, e.g. eqs (16)-(22) of \cite{Ruggiero:2002}. Furthermore (on
certain special conditions, see section \ref{sec:Linearized-Gravity}) \linebreak
geodesics, precession and forces on gyroscopes are described in terms
of these fields in a \linebreak form similar to their electromagnetic counterparts,
e.g. \cite{Ruggiero:2002}, \linebreak \cite{Gravitation and Inertia}. Such analogy may
actually be cast in an exact form using the 3+1 splitting of spacetime
(see \cite{Jantzen,Natario}).

These are analogies comparing physical quantities (electromagnetic
forces) from one theory with inertial gravitational forces (i.e. fictitious
forces, that can be gauged away by moving to a freely falling frame,
due to the equivalence principle); it is clear that (non-spinning)
test particles in a gravitational field move with zero acceleration
$DU^{\alpha}/d\tau=0$; and that the spin 4-vector of a gyroscope
undergoes Fermi-Walker transport $DS^{\alpha}/d\tau=S_{\sigma}U^{\alpha}DU^{\sigma}/d\tau$,
with no real torques applied on it. In this sense the gravito-electromagnetic
fields are pure coordinate artifacts, attached to the observer's frame.

However, these approaches describe also (not through the {}``gravito-electromagnetic''
fields themselves, but through their derivatives; and, again, under
very special conditions) tidal effects, like the force applied on
a gyroscope. And these are covariant effects, implying physical gravitational
forces. 

Herein we will discuss under which precise conditions a similarity
between gravity and electromagnetism occurs (that is, under which
conditions the physical analogy $\bar{h}_{0\mu}\leftrightarrow A_{\mu}$
holds, and Eqs. like (16)-(22) of \cite{Ruggiero:2002} have a physical
content). For that we will make use of the tidal tensor formalism
introduced in \cite{CostaHerdeiro2008}. The advantage of this formalism
is that, by contrast with the approaches mentioned above, it is based
on quantities which can be covariantly defined in both theories ---
tidal forces (the only physical forces present in gravity) --- which
allows for a more transparent comparison between the electromagnetic
(EM) and gravitational (GR) interactions.

\begin{table}[h]
\caption{\label{analogy}The gravito-electromagnetic analogy based on tidal
tensors. }

\begin{centering}
\setlength{\arrayrulewidth}{0.8pt}\begin{tabular}{>{\centering}p{39.6ex}c>{\centering}p{41.9ex}c}
\hline 
\multicolumn{2}{c}{\raisebox{3ex}{}\raisebox{0.5ex}{Electromagnetism}} & \multicolumn{2}{c}{\raisebox{3ex}{}\raisebox{0.5ex}{Gravity}}\tabularnewline
\hline 
\raisebox{3ex}{}Worldline deviation:  &  & \raisebox{3ex}{}Geodesic deviation:  & \tabularnewline
\raisebox{6ex}{}\raisebox{2ex}{${\displaystyle \frac{D^{2}\delta x^{\alpha}}{d\tau^{2}}=\frac{q}{m}E_{\,\,\,\beta}^{\alpha}\delta x^{\beta}},\,\,\, E_{\,\,\,\beta}^{\alpha}\equiv F_{\ \mu;\beta}^{\alpha}U^{\mu}$}  & \raisebox{2ex}{(1a) \,\,}  & \raisebox{5.5ex}{}\raisebox{2ex}{${\displaystyle \frac{D^{2}\delta x^{\alpha}}{d\tau^{2}}=-\mathbb{E}_{\,\,\,\beta}^{\alpha}\delta x^{\beta}},\,\,\,\mathbb{E}_{\,\,\,\beta}^{\alpha}\equiv R_{\,\,\,\mu\beta\nu}^{\alpha}U^{\mu}U^{\nu}$}  & \raisebox{2ex}{(1b)}\tabularnewline
\hline 
\raisebox{3ex}{}Force on magnetic dipole:  &  & \raisebox{3ex}{}Force on gyroscope:  & \tabularnewline
\raisebox{6ex}{}\raisebox{2ex}{${\displaystyle F_{EM}^{\beta}=\frac{q}{2m}B_{\alpha}^{\,\,\,\beta}S^{\alpha}},\,\,\, B_{\,\,\,\beta}^{\alpha}\equiv\star F_{\ \mu;\beta}^{\alpha}U^{\mu}$}  & \raisebox{2ex}{(2a)}  & \raisebox{6ex}{}\raisebox{2ex}{~~${\displaystyle F_{G}^{\beta}=-\mathbb{H}_{\alpha}^{\,\,\,\beta}S^{\alpha}},\,\,\,\mathbb{H}_{\,\,\,\beta}^{\alpha}\equiv\star R_{\,\,\,\mu\beta\nu}^{\alpha}U^{\mu}U^{\nu}$}  & \raisebox{2ex}{(2b)}\tabularnewline
\hline 
\raisebox{3ex}{}Maxwell Equations:  &  & \raisebox{3ex}{}Eqs. Grav. Tidal Tensors:  & \tabularnewline
\raisebox{3ex}{}$E_{\,\,\,\alpha}^{\alpha}=4\pi\rho_{c}$  & (3a)  & \raisebox{3ex}{}$\mathbb{E}_{\,\,\,\alpha}^{\alpha}=4\pi\left(2\rho_{m}+T_{\,\,\alpha}^{\alpha}\right)$  & (3b)\tabularnewline
\raisebox{3.5ex}{}$E_{[\alpha\beta]}=\frac{1}{2}F_{\alpha\beta;\gamma}U^{\gamma}$  & (4a)  & \raisebox{3.5ex}{}$\mathbb{E}_{[\alpha\beta]}=0$  & (4b)\tabularnewline
\raisebox{3.5ex}{}$B_{\,\,\,\alpha}^{\alpha}=0$  & (5a)  & \raisebox{3.5ex}{}$\mathbb{H}_{\,\,\,\alpha}^{\alpha}=0$  & (5b)\tabularnewline
\raisebox{5.5ex}{}\raisebox{2ex}{$B_{[\alpha\beta]}=\frac{1}{2}\star F_{\alpha\beta;\gamma}U^{\gamma}-2\pi\epsilon_{\alpha\beta\sigma\gamma}j^{\sigma}U^{\gamma}$}  & \raisebox{2ex}{(6a)}  & ~~\raisebox{5.5ex}{}\raisebox{2ex}{$\mathbb{H}_{[\alpha\beta]}=-4\pi\epsilon_{\alpha\beta\sigma\gamma}J^{\sigma}U^{\gamma}$}~~  & \raisebox{2ex}{(6b)}\tabularnewline
\hline
\end{tabular}\\
 
\par\end{centering}

\begin{raggedright}
{\scriptsize $\rho_{c}=-j^{\alpha}U_{\alpha}$ and $j^{\alpha}$ are,
respectively, the charge density and current 4-vector; $\rho_{m}=T_{\alpha\beta}U^{\alpha}U^{\beta}$
and $J^{\alpha}=-T_{\,\beta}^{\alpha}U^{\beta}$ are the mass/energy
density and current (quantities measured by the observer of 4-velocity
$U^{\alpha}$); $T_{\alpha\beta}\equiv$ energy-momentum tensor; $S^{\alpha}\equiv$
spin 4-vector; $\star\equiv$ Hodge dual. We use $\tilde{e}_{0123}=-1$.}
\par\end{raggedright}
\end{table}

The tidal tensor formalism unveils a new gravito-electromagnetic analogy,
summarized in Table \ref{analogy}, based on exact and covariant equations.
These equations make clear key differences, and under which conditions
a similarity between the two interactions may occur.

Eqs. (1) are the worldline deviation equations yielding the relative
acceleration of two neighboring particles (connected by the infinitesimal
vector $\delta x^{\alpha}$) with the \textit{same} 4-velocity $U^{\alpha}$
(and the same $q/m$ ratio, in the electromagnetic case). These equations
manifest the physical analogy between electric tidal tensors: $\mathbb{E}_{\alpha\beta}\leftrightarrow E_{\alpha\beta}$.

Eq. (2a) yields the electromagnetic force exerted on a magnetic dipole
moving with 4-velocity $U^{\alpha}$, and is the covariant generalization
of the usual 3-D expression $\mathbf{F_{EM}}=\nabla(\mathbf{S}.\mathbf{B})q/2m$
(valid only in the dipole's proper frame); Eq. (2b) is exactly the
Papapetrou-Pirani equation for the gravitational force exerted on
a spinning test particle. In both (2a) and (2b), Pirani's supplementary
condition $S_{\mu\nu}U^{\nu}=0$ is assumed (c.f. \cite{CostaHerdeiro2009}).
These equations manifest the physical analogy between magnetic tidal
tensors: $B_{\alpha\beta}\leftrightarrow\mathbb{H}_{\alpha\beta}$.

Taking the traces and antisymmetric parts of the EM tidal tensors,
one obtains Eqs. (3a)-(6a), which are explicitly covariant forms for
each of Maxwell equations. Eqs. (3a) and (6a) are, respectively, the
time and space projections of Maxwell equations $F_{\ \ \ ;\beta}^{\alpha\beta}=4\pi j^{\alpha}$;
i.e., they are, respectively, covariant forms of $\nabla\cdot\mathbf{E}=4\pi\rho_{c}$
and $\nabla\times\mathbf{B}=\partial\mathbf{E}/\partial t+4\pi\mathbf{j}$;
Eqs. (4a) and (5a) are the space and time projections of the electromagnetic
Bianchi identity $\star F_{\ \ \ ;\beta}^{\alpha\beta}=0$; i.e.,
they are covariant forms for $\nabla\times\mathbf{E}=-\partial\mathbf{B}/\partial t$
and $\nabla\cdot\mathbf{B}=0$. These equations involve only tidal
tensors and sources, which can be seen substituting the following
decomposition (or its Hodge dual) in (4a) and (6a): \begin{equation}
F_{\alpha\beta;\gamma}=2U_{[\alpha}E_{\beta]\gamma}+\epsilon_{\alpha\beta\mu\sigma}B_{\,\,\,\gamma}^{\mu}U^{\sigma}\
.\label{Fdecomp}\end{equation}
It is then straightforward to obtain the \emph{physical} gravitational
analogues of Maxwell equations: one just has to apply the same procedure
to the gravitational tidal tensors, i.e., write the equations for
their traces and antisymmetric parts (that is more easily done decomposing
the Riemann tensor in terms of the Weyl tensor and source terms, see
\cite{CostaHerdeiro2007} sec. 2), which leads to Eqs. (3b) - (6b).
Underlining the analogy with the situation in electromagnetism, Eqs.
(3b) and (6b) turn out to be the time-time and and time-space projections
of Einstein equations $R_{\mu\nu}=8\pi(T_{\mu\nu}-\frac{1}{2}g_{\mu\nu}T_{\,\,\,\alpha}^{\alpha})$,
and Eqs. (4b) and (5b) the time-space and time-time projections of
the algebraic Bianchi identities $\star R_{\ \ \ \gamma\beta}^{\gamma\alpha}=0$.

\subsection{Gravity vs Electromagnetism}

\emph{Charges} --- the gravitational analogue of $\rho_{c}$ is $2\rho_{m}+T_{\,\,\alpha}^{\alpha}$
($\rho_{m}+3p$ for a perfect fluid) $\Rightarrow$ in gravity, pressure
and all material stresses contribute as sources.

\emph{Ampere law} --- in stationary (in the observer's rest frame)
setups, $\star F_{\alpha\beta;\gamma}U^{\gamma}$ vanishes and equations
(6a) and (6b) match up to a factor of 2 $\Rightarrow$ currents of
mass/energy source gravitomagnetism like currents of charge source
magnetism. 

\emph{Symmetries of Tidal Tensors} --- The GR and EM tidal tensors
do not generically exhibit the same symmetries, signaling fundamental
differences between the two interactions. In the general case of fields
that are time dependent in the observer's rest frame (that is the
case of an intrinsically non-stationary field, or an observer moving
in a stationary field), the electric tidal tensor $E_{\alpha\beta}$
possesses an antisymmetric part, which is the covariant derivative
of the Maxwell tensor along the observer's worldline; there is also
an antisymmetric contribution $\star F_{\alpha\beta;\gamma}U^{\gamma}$
to $B_{\alpha\beta}$. These terms consist of time projections of
EM tidal tensors (cf. decomposition \ref{Fdecomp}), and contain the
laws of electromagnetic induction. The gravitational tidal tensors,
by contrast, are symmetric (in vacuum, in the magnetic case) and spatial,
manifesting the absence of analogous effects in gravity.

\emph{Gyroscope vs. magnetic dipole} --- According to Eqs. (2), both
in the case of the magnetic dipole and in the case of the gyroscope,
it is the magnetic tidal tensor, \emph{as seen by the test particle}
($U^{\alpha}$ in Eqs. (2) is the gyroscope/dipole 4-velocity), that
determines the force exerted upon it. Hence, from Eqs. (6), we see
that the forces can be similar only if the fields are stationary (besides
weak) in the gyroscope/dipole frame, i.e., when it is at {}``rest''
in a stationary field. Eqs. (2) also tell us that in gravity the angular
momentum $S$ plays the role of the magnetic moment $\mu=S(q/2m)$;
the relative minus sign manifests that masses/charges of the same
sign attract/repel one another in gravity/electromagnetism, as do
charge/mass currents with parallel velocity.

\section{Linearized Gravity\label{sec:Linearized-Gravity}}

If the fields are stationary in the observer's rest frame, the GR
and EM tidal tensors have the same symmetries, which by itself does
not mean a close similarity between the two interactions (note that
despite the analogy in Table 1, EM tidal tensors are linear, whereas
the GR ones are not). But in two special cases a matching between
tidal tensors occurs: ultrastationary spacetimes (where the gravito-magnetic
tidal tensor is linear, see \cite{CostaHerdeiro2008} Sec. IV) and
linearized gravitational perturbations, which is the case of interest
for astronomical applications. 

Consider an arbitrary electromagnetic field $A^{\alpha}=(\phi,\mathbf{A})$
and arbitrary perturbations around Minkowski spacetime in the form%
\footnote{In the previous sections we were putting $c=1$. In this section we
re-introduce the speed of light in order to facilitate comparison
with relevant literature.%
}

\begin{equation}
ds^{2}=-c^{2}\left(1-2\frac{\Phi}{c^{2}}\right)dt^{2}-\frac{4}{c}\mathcal{A}_{j}dtdx^{j}+\left[\delta_{ij}+2\frac{\Theta_{ij}}{c^{2}}\right]dx^{i}dx^{j}\ .\label{Linear pert}\end{equation}

\textbf{Tidal effects.} --- The GR and EM tidal tensors from these
setups will be in general very different, as is clear from equations
(3-6), and as one may check from the explicit expressions in \cite{CostaHerdeiro2008}.

But if one considers time independent fields, and a static observer
of 4-velocity $U^{\mu}=c\delta_{0}^{\mu}$, then the \emph{linearized}
gravitational tidal tensors match their electromagnetic counterparts
identifying $(\phi,A^{i})\leftrightarrow(\Phi,\mathcal{A}^{i})$ (in
expressions below colon represents partial derivatives; $\epsilon_{ijk}\equiv$
Levi Civita symbol): \begin{equation}
\mathbb{E}_{ij}\simeq-\Phi_{,ij}\stackrel{\Phi\leftrightarrow\phi}{=}E_{ij},\quad\mathbb{H}_{ij}\simeq\epsilon_{i}^{\,\,\, lk}\mathcal{A}_{k,lj}\stackrel{\mathcal{A}\leftrightarrow A}{=}B_{ij}\ .\label{MatchingLinear}\end{equation}
 This suggests the physical analogy $(\phi,A^{i})\leftrightarrow(\Phi,\mathcal{A}^{i})$,
and defining the {}``gravito-electro-magnetic fields'' $\mathbf{E_{G}}=-\nabla\Phi$
and $\mathbf{B_{G}}=\nabla\times\bm{\mathcal{A}}$, in analogy with
the electromagnetic fields $\mathbf{E}=-\nabla\phi,$ $\mathbf{B}=\nabla\times\mathbf{A}$.
In terms of these fields we have $\mathbb{E}_{ij}\simeq(E_{G})_{i,j}$
and $\mathbb{H}_{ij}\simeq(B_{G})_{i,j}$, in analogy with the electromagnetic
tidal tensors $E_{ij}=E_{i,j}$ and $B_{ij}=B_{i,j}$. 

The matching (\ref{MatchingLinear}) means that a gyroscope at rest
(relative to the static observer) will feel a force $F_{G}^{\alpha}$
similar to the electromagnetic force $F_{EM}^{\alpha}$ on a magnetic
dipole, which in this case take the very simple forms (time components
are zero): \begin{equation}
\mathbf{F_{EM}}=\frac{q}{2mc}\nabla(\mathbf{B}.\mathbf{S});\ \ \ \ \ F_{G}^{j}=-\frac{1}{c}\mathbb{H}^{ij}S_{i}\approx-\frac{1}{c}(B_{G})^{i,j}S_{i}\ \Leftrightarrow\ \mathbf{F_{G}}=-\frac{1}{c}\nabla(\mathbf{B_{G}}.\mathbf{S})\ .\label{FG_Stationary}\end{equation}
Had we considered gyroscopes/dipoles with different 4-velocities,
not only the expressions for the forces would be more complicated,
but also the gravitational force would significantly differ from the
electromagnetic one, as one may check comparing Eqs. (12) with (17)-(20)
of \cite{CostaHerdeiro2008}. This will be exemplified in section
\ref{Translational-vs.-Rotational}.

The matching (\ref{MatchingLinear}) also means, by similar arguments,
that the relative acceleration between two neighboring masses $D^{2}\delta x^{i}/d\tau^{2}=-\mathbb{E}^{ij}\delta x_{j}$
is similar to the relative acceleration between two charges (with
the same $q/m$): $D^{2}\delta x^{i}/d\tau^{2}=E^{ij}\delta x_{j}(q/m)$,
\emph{at the instant} when the test particles have 4-velocity $U^{\alpha}=c\delta_{0}^{\alpha}$
(i.e., are \emph{at rest} relative to the static observer $\mathcal{O}$).

\textbf{Gyroscope precession.} --- The evolution of the spin vector
of the gyroscope is given by the Fermi-Walker transport law, which,
for a gyroscope at rest reads $DS^{i}/d\tau=0$; hence, we have, in
the coordinate basis, Eq. (\ref{PrecessGen}a). The last term of Eq.
(\ref{PrecessGen}a) vanishes if we express $\mathbf{S}$ in the local
orthonormal tetrad $e^{\hat{\alpha}}$: $S^{i}=S^{\hat{i}}e_{\,\hat{i}}^{i}$,
where to linear order $e_{\,\hat{i}}^{i}=\delta_{\ \hat{i}}^{i}-\Theta_{\ \ \hat{i}}^{i}/c^{2}$;
in this fashion we obtain Eq. (\ref{PrecessGen}b), which is similar
to the precession of a magnetic dipole in a magnetic field $d\mathbf{S}/dt=q\mathbf{S}\times\mathbf{B}/2mc$:
\begin{equation}
\frac{dS^{i}}{dt}=-c\Gamma_{0j}^{i}S^{j}=-\frac{1}{c}\left[(\mathbf{S}\times\mathbf{B_{G}})^{i}+\frac{1}{c}\frac{\partial\Theta_{\ }^{ij}}{\partial t}S_{j}\right]\ \ (a);\ \ \ \ \ \ \frac{dS^{\hat{i}}}{dt}=-\frac{1}{c}(\mathbf{S}\times\mathbf{B_{G}})^{\hat{i}}\ \ (b).\label{PrecessGen}\end{equation}
 Thus, in the special case of gyroscope precession, the linear gravito-electromagnetic
analogy holds even if the fields vary with time.

\textbf{Geodesics.} --- The space part of the equation of geodesics
$U_{\ ,\beta}^{\alpha}U^{\beta}=-\Gamma_{\beta\gamma}^{\alpha}U^{\beta}U^{\gamma}$
is given, to first order in the perturbations and in test particle's
velocity, by ($a^{i}\equiv d^{2}x^{i}/dt^{2})$:\begin{eqnarray}
\mathbf{a} & = & \nabla\Phi+\frac{2}{c}\frac{\partial\bm{\mathcal{A}}}{\partial t}-\frac{2}{c}\mathbf{v}\times(\nabla\times\bm{\mathcal{A}})-\frac{1}{c^{2}}\left[\frac{\partial\Phi}{\partial t}\mathbf{v}+2\frac{\partial\Theta_{\ j}^{i}}{\partial t}v^{j}\mathbf{e_{i}}\right]\ .\label{geoGeneral}\end{eqnarray}
Comparing with the electromagnetic Lorentz force:\begin{equation}
\mathbf{a}=\frac{q}{m}\left[-\nabla\phi-\frac{1}{c}\frac{\partial\mathbf{A}}{\partial t}+\frac{\mathbf{v}}{c}\times(\nabla\times\mathbf{A})\right]=\frac{q}{m}\left[\mathbf{E}+\frac{\mathbf{v}}{c}\times\mathbf{B}\right]\,,\label{Lorentz}\end{equation}
these equations do not manifest, in general, a close analogy. Note
that the last term of (\ref{geoGeneral}), which has no electromagnetic
analogue, is, for the problem at hand (see next section), of the same
order of magnitude as the second and third terms. But when one considers
stationary fields, then (\ref{geoGeneral}) takes the form $\mathbf{a}=-\mathbf{E_{G}}-2\mathbf{v}\times\mathbf{B_{G}}/c$
analogous to (\ref{Lorentz}).

Note the difference between this analogy and the one from the tidal
effects considered above: in the case of the latter, the similarity
occurs only when the \emph{test particle} sees time independent \emph{fields}
(fields $\equiv$ derivatives of potentials/of metric perturbations);
for geodesics, it is when \emph{the observer} (not the test particle!)
sees a time independent \emph{potential }($\phi$)\emph{/metric perturbations}($\Phi,\Theta_{ij}$).

\subsection{\label{Translational-vs.-Rotational}Translational vs. Rotational
Mass Currents}

The existence of a similarity between gravity and electromagnetism
thus relies on the time dependence of the mass currents: if the currents
are (nearly) stationary, for instance from a spinning celestial body,
the gravitational field generated is analogous to a magnetic field;
an example is the gravitomagnetic field due to the rotation of the Earth,
detected on LAGEOS data by \cite{Ciufolini Lageos} (and which is
also the subject of experimental scrutiny by the Gravity Probe B and
the upcoming LARES missions). But when the currents seen by the observer
vary with time --- e.g. the ones resulting from translation of the
celestial body, considered in \cite{SoffelKlioner} --- then the dynamics
differ significantly.

\begin{figure}[h]

\begin{centering}
\includegraphics[width=123.2mm]{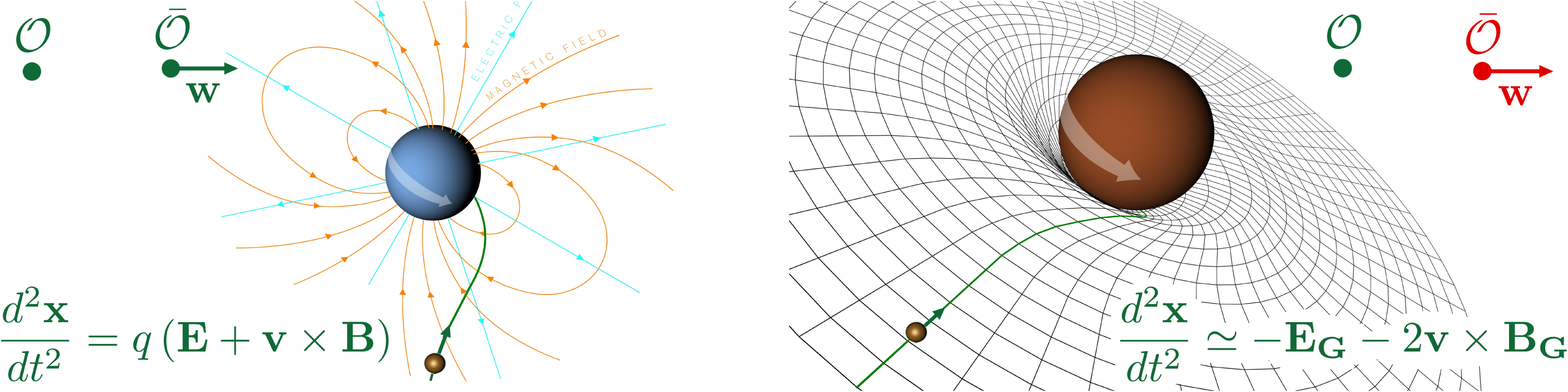} 

\par\end{centering}

\caption{Spinning charge vs. spinning mass }

\begin{centering}
\label{fig1} 
\par\end{centering}
\end{figure}

\textbf{Rotational Currents.} --- We will start by the well known
analogy between the electromagnetic field of a spinning charge (charge
$Q$, magnetic moment $\mu$) and the gravitational field (in the
far region $r\rightarrow\infty$) of a rotating celestial body (mass
$m$, angular momentum $J$), see Fig.\,\ref{fig1}

The electromagnetic field of the spinning charge is described by the
4-potential $A^{\alpha}=(\phi,\mathbf{A})$, given by (\ref{GravPot}a).
The spacetime around the spinning mass is asymptotically described
by the linearized Kerr solution, obtained by putting in (\ref{Linear pert})
the perturbations (\ref{GravPot}b) :\begin{equation}
\phi=\frac{Q}{r}\ ,\ \ \ \mathbf{A}=\frac{1}{c}\frac{\bm{\mathbf{\mu}}\times\mathbf{r}}{r^{3}}\ \ \ (a);\ \ \ \ \ \ \Phi=\frac{M}{r}\ ,\ \ \ \bm{\mathcal{A}}=\frac{1}{c}\frac{\mathbf{J}\times\mathbf{r}}{r^{3}}\ ,\ \ \Theta_{ij}=\Phi\delta_{ij}\ \ \ (b).\ \ \ \label{GravPot}\end{equation}
For the observer at rest $\mathcal{O}$ the gravitational tidal tensors
asymptotically match the electromagnetic ones, identifying the appropriate
parameters:{\small \[
\mathbb{E}_{ij}\simeq\frac{M}{r^{3}}\delta_{ij}-\frac{3Mr_{i}r_{j}}{r^{5}}\stackrel{M\leftrightarrow Q}{=}E_{ij};\ \ \ \ \mathbb{H}_{ij}\simeq\frac{3}{c}\left[\frac{(\mathbf{r}.\mathbf{J})}{r^{5}}\delta_{ij}+2\frac{r_{(i}J_{j)}}{r^{5}}-5\frac{(\mathbf{r}.\mathbf{J})r_{i}r_{j}}{r^{7}}\right]\stackrel{J\leftrightarrow\mu}{=}B_{ij}\]
}(all the time components are zero for this observer). This me{\small ans
that} $\mathcal{O}$ will find a similarity between \emph{physical}
(i.e., tidal) gravitational forces and their electromagnetic counterparts:
the gravitational force $F_{G}^{i}=-\mathbb{H}^{ji}S_{j}/c$ exerted
on a gyroscope carried by $\mathcal{O}$ is similar to the force $F_{EM}^{i}=qB^{ji}S_{j}/2mc$
on a magnetic dipole; and the worldline deviation $D^{2}\delta x^{i}/d\tau^{2}=-\mathbb{E}^{ij}\delta x_{i}$
of two masses dropped from rest is similar to the deviation between
two charged particles with the same $q/m$.

Moreover, observer $\mathcal{O}$ will see test particles moving on
geodesics described by equations analogous to the electromagnetic
Lorentz force (see Fig. \ref{fig1}). 

\textbf{Translational Currents.} --- For the observer $\bar{\mathcal{O}}$
moving with velocity $\mathbf{w}$ relative to the mass/charge of
Fig. \ref{fig1}, however, the electromagnetic and gravitational interactions
will look significantly different. For simplicity we will specialize
here to the case where $\mathbf{J}=\bm{\mu}=0$, so that the mass/charge
currents seen by $\bar{\mathcal{O}}$ arise solely from translation.
To obtain the electromagnetic 4-potential $A^{\bar{\alpha}}$ in the
frame $\bar{\mathcal{O}}$, we apply the boost $A^{\bar{\alpha}}=\Lambda_{\ \alpha}^{\bar{\alpha}}A^{\alpha}=(\bar{\phi},\bar{\mathbf{A}})$,
where $\Lambda_{\ \alpha}^{\bar{\alpha}}\equiv\partial\bar{x}^{\bar{\alpha}}/\partial x^{\alpha}$,
using the expansion of Lorentz transformation (as done in e.g. \cite{WillNordvedt1972}):
\begin{equation}
t=\bar{t}\left(1+\frac{w^{2}}{2c^{2}}+\frac{3w^{4}}{8c^{4}}\right)+\left(1+\frac{w^{2}}{2c^{2}}\right)\frac{\mathbf{\bar{x}}.\mathbf{w}}{c^{2}};\ \ \ \ \mathbf{x}=\mathbf{\bar{x}}+\frac{1}{2c^{2}}(\bar{\mathbf{x}}.\mathbf{w})\mathbf{w}+\left(1+\frac{w^{2}}{2c^{2}}\right)\mathbf{w}\bar{t}\ ,\label{Boost}\end{equation}
yielding, to order $c^{-2}$, $A^{\bar{\alpha}}=(\bar{\phi},\mathbf{\bar{A}})$,
with $\bar{\phi}=Q(1+w^{2}/2c^{2})/r$ and $\mathbf{\bar{A}}=-Q\mathbf{w}/rc$.
To obtain $A^{\bar{\alpha}}$ in the coordinates ($\bar{x}^{i},\bar{t}$)
of $\bar{\mathcal{O}}$, we must also express $r$ (which denotes
the distance between the source and the point of observation, in the
frame $\mathcal{O}$) in terms of $R\equiv|\mathbf{\bar{r}}+\mathbf{w}\bar{t}|$,
i.e., the distance between the source and the point of observation
in the frame $\bar{\mathcal{O}}$. Using transformation (\ref{Boost}),
we obtain: $r^{-1}=R^{-1}[1-(\mathbf{w}.\mathbf{R})^{2}/(2R^{2}c^{2})]$,
and finally the electromagnetic potentials seen by $\bar{\mathcal{O}}$:\begin{equation}
\bar{\phi}=\frac{Q}{R}\left(1+\frac{w^{2}}{2c^{2}}-\frac{(\mathbf{w}.\mathbf{R})^{2}}{4R^{2}c^{2}}\right);\ \ \ \ \ \ \mathbf{\bar{A}}=-\frac{1}{c}\frac{Q}{R}\mathbf{w}\ .\label{EMF Obar}\end{equation}
The metric of the spacetime around a point mass, in the coordinates
of $\bar{\mathcal{O}}$, is also obtained using transformation (\ref{Boost}),
which is accurate to Post Newtonian order, by an analogous procedure.
First we apply the Lorentz boost $g_{\bar{\alpha}\bar{\beta}}=\Lambda_{\ \bar{\alpha}}^{\alpha}\Lambda_{\ \bar{\beta}}^{\beta}g_{\alpha\beta}$
to the metric (\ref{GravPot}) (with $\mathcal{A}=0$); then, expressing
$r$ in terms of $R$, we finally obtain (note that, although we are
not putting the bars therein, indices $\alpha=0,i$ in the following
expressions refer to the coordinates of $\bar{\mathcal{O}}$): \begin{eqnarray}
g_{00} & = & -1+2\frac{M}{Rc^{2}}+\frac{4Mw^{2}}{Rc^{4}}-\frac{M(\mathbf{w}.\mathbf{R})^{2}}{c^{4}R^{3}}\equiv-1+\frac{2\bar{\Phi}}{c^{2}};\ \ \nonumber \\
\ g_{0i} & = & \frac{4Mw_{i}}{Rc^{3}}\equiv-\frac{2\bar{\mathcal{A}}_{i}}{c^{2}};\ \ \ \ \ \ \ \ \ \ g_{ij}=\left[1+2\frac{M}{Rc^{2}}\right]\delta_{ij}\equiv\left[1+2\frac{\bar{\Theta}}{c^{2}}\right]\delta_{ij}\ ,\label{GObar}\end{eqnarray}
where we retained terms up to $c^{-4}$ in $g_{00}$, up to $c^{-3}$
in $g_{i0}$ and $c^{-2}$ in $g_{ij}$, as usual in Post-Newtonian
approximation. This matches, to linear order in $M$, Eqs. (5) of
\cite{SoffelKlioner} for the case of one single source; or e.g. Eqs.
(11) of \cite{Nordvedt1988} (in the case of the latter, an additional
gauge choice, Eq. (19) of \cite{WillNordvedt1972}, was made). The
metric (\ref{GObar}), like the electromagnetic potential (\ref{EMF Obar}),
is now time dependent, since $\mathbf{R}(\bar{t})=\mathbf{\bar{r}}+\mathbf{w}\bar{t}$.

The gravitational tidal tensors seen by $\bar{\mathcal{O}}$ are ({\small $\mathbb{E}_{\alpha0}=\mathbb{E}_{0\alpha}=\mathbb{H}_{\alpha0}=\mathbb{H}_{0\alpha}=0$}):{\small \begin{eqnarray}
\mathbb{E}_{ij} & = & -\bar{\Phi}_{,ij}-\frac{2}{c}\frac{\partial}{\partial\bar{t}}\bar{\mathcal{A}}_{(i,j)}-\frac{1}{c^{2}}\frac{\partial^{2}}{\partial\bar{t}^{2}}\bar{\Theta}\delta_{ij}\nonumber \\
 & = & \frac{M\delta_{ij}}{R^{3}}\left[1+\frac{3w^{2}}{c^{2}}-\frac{9}{2}\frac{(\mathbf{R}.\mathbf{w})^{2}}{c^{2}R^{2}}\right]-\frac{3MR_{i}R_{j}}{R^{5}}\left[1+\frac{2w^{2}}{c^{2}}-\frac{5(\mathbf{R}.\mathbf{w})^{2}}{2c^{2}R^{2}}\right]\nonumber \\
 &  & -\frac{3Mw_{i}w_{j}}{c^{2}R^{3}}+\frac{6Mw_{(i}R_{j)}(\mathbf{R}.\mathbf{w})}{c^{2}R^{5}};\label{Egij}\\
\mathbb{H}_{ij} & = & \epsilon_{i}^{\,\,\, lk}\bar{\mathcal{A}}_{k,lj}-\frac{1}{c}\epsilon_{ij}^{\ \ l}\frac{\partial\bar{\Theta}_{,l}}{\partial\bar{t}}\ =\frac{M}{cR^{3}}\left[3\epsilon_{ij}^{\ \ k}w_{k}-\frac{3}{R^{2}}(\mathbf{R}.\mathbf{w})\epsilon_{ij}^{\ \ k}R_{k}-\frac{6}{R^{2}}(\mathbf{R}\times\mathbf{w})_{i}R_{j}\right],\label{Hij}\end{eqnarray}
}which significantly differ from the electromagnetic ones ({\small $E_{0\alpha}=B_{0\alpha}=0$}):{\small \begin{eqnarray}
E_{ij} & = & -\bar{\phi}_{,ij}-\frac{1}{c}\frac{\partial}{\partial\bar{t}}\bar{A}_{i;j}\ =\ E_{i,j}\nonumber \\
 & = & \frac{Q\delta_{ij}}{R^{3}}\left[1+\frac{w^{2}}{2c^{2}}-\frac{3}{4}\frac{(\mathbf{R}.\mathbf{w})^{2}}{c^{2}R^{2}}\right]-\frac{3QR_{i}R_{j}}{R^{5}}\left[1+\frac{w^{2}}{2c^{2}}-\frac{5(\mathbf{R}.\mathbf{w})^{2}}{4c^{2}R^{2}}\right]\nonumber \\
 &  & -\frac{Qw_{i}w_{j}}{2c^{2}R^{3}}+\frac{3Qw_{[i}R_{j]}(\mathbf{R}.\mathbf{w})}{c^{2}R^{5}};\label{Eij}\\
E_{i0} & = & -\frac{1}{c}\frac{\partial}{\partial\bar{t}}\bar{\phi}_{;i}-\frac{1}{c^{2}}\frac{\partial^{2}\bar{A}_{i}}{\partial\bar{t}^{2}}\ \equiv\frac{1}{c}\ \frac{\partial E_{i}}{\partial\bar{t}}\ =\ \frac{Q}{cR^{3}}\left[w_{i}-\frac{3(\mathbf{R}.\mathbf{w})R_{i}}{R^{2}}\right];\label{Ei0}\\
B_{ij} & = & \epsilon_{i}^{\ lm}\bar{A}_{m;lj}\ \equiv\ B_{i,j}=\frac{Q}{cR^{3}}\left[\epsilon_{ij}^{\ \ k}w_{k}-\frac{3}{R^{2}}(\mathbf{R}\times\mathbf{w})_{i}R_{j}\right];\label{Bij}\\
B_{i0}\  & = & \frac{1}{c}\frac{\partial B_{i}}{\partial\bar{t}}=-\ \frac{3Q}{c^{2}R^{5}}(\mathbf{R}.\mathbf{w})(\mathbf{R}\times\mathbf{w})_{i}\ .\label{Bi0}\end{eqnarray}
}Note in particular that, unlike their gravitational counterparts,
$E_{\alpha\beta}$ and $B_{\alpha\beta}$ are not symmetric, and have
non-zero time components. The antisymmetric parts $E_{[ij]}=E_{[i,j]}$
and $B_{[ij]}=B_{[i,j]}$ above are (vacuum) Maxwell equations $\nabla\times\mathbf{E}=-(1/c)\partial\mathbf{B}/\partial t$
and $\nabla\times\mathbf{B}=(1/c)\partial\mathbf{E}/\partial t$,
implying that a time varying electric/magnetic field endows the magnetic/electric
tidal tensor with an antisymmetric part. For instance, a time varying
electric field will always induce a force on a magnetic dipole. The
fact that $\mathbb{E}_{\alpha\beta}$ and $\mathbb{H}_{\alpha\beta}$
are symmetric reflects the absence of analogous gravitational effects.
The time component $B_{i0}$ means that the force on a magnetic dipole
(magnetic moment $\mu=q/2m$) will have a time component $(F_{EM})_{0}=(1/c)\bm{\mu}.\partial\mathbf{B}/\partial t$,
which (see \cite{CostaHerdeiro2009} sec. 1.2) is minus the power
transferred to the dipole by Faraday's law of induction (and is reflected
in the variation of the dipole's proper mass $m=-P^{\alpha}U_{\alpha}/c^{2}$).
Again, this is an effect which has no gravitational counterpart: $\mathbb{H}_{\alpha0}=\mathbb{H}_{0\alpha}=0$,
thus $(F_{G})_{0}=0$, and the proper mass of the gyroscope is a constant
of the motion. 

The space part of the geodesic equation for a test particle of velocity
$\mathbf{v}$ is:{\small \begin{eqnarray}
\mathbf{a} & = & \nabla\bar{\Phi}+\frac{2}{c}\frac{\partial\bm{\bar{\mathcal{A}}}}{\partial\bar{t}}-2\mathbf{v}\times(\nabla\times\bm{\bar{\mathcal{A}}})-\frac{3}{c^{2}}\frac{\partial}{\partial\bar{t}}\left(\frac{M}{R}\right)\mathbf{v}\label{GeoTrans}\\
 & = & -\frac{M}{R^{3}}\left[1+\frac{2w^{2}}{c^{2}}-\frac{3(\mathbf{R}.\mathbf{w})^{2}}{2c^{2}R^{2}}\right]\mathbf{R}+\frac{3M(\mathbf{R}.\mathbf{w})}{c^{2}R^{3}}\mathbf{w}-\frac{4M}{c^{2}R^{3}}\mathbf{v}\times(\mathbf{R}\times\mathbf{w})+\frac{3}{c^{2}}\frac{M}{R^{3}}(\mathbf{R}.\mathbf{w})\mathbf{v}\ ,\nonumber \end{eqnarray}
}which matches equation (10) of \cite{SoffelKlioner}, or (7) of \cite{Nordvedt1973},
again, in the special case of only one source, and keeping therein
only linear terms in the perturbations and test particle's velocity
$\mathbf{v}$. 

Comparing with its electromagnetic counterpart{\small \[
\left(\frac{m}{q}\right)\mathbf{a}=\mathbf{E}+\frac{\mathbf{v}}{c}\times\mathbf{B}=\frac{Q}{R^{3}}\left[1+\frac{w^{2}}{2c^{2}}-\frac{3(\mathbf{R}.\mathbf{w})^{2}}{4c^{2}R^{2}}\right]\mathbf{R}-\frac{1}{2}\frac{Q(\mathbf{R}.\mathbf{w})}{c^{2}R^{3}}\mathbf{w}+\frac{Q}{c^{2}R^{3}}\mathbf{v}\times(\mathbf{R}\times\mathbf{w})\]
}we find them similar to a certain degree (up to some factors), except
for the last term of (\ref{GeoTrans}). That term signals a difference
between the two interactions, because it means that there is a velocity
dependent acceleration which is parallel to the velocity; that is
in contrast with the situation in electromagnetism, where the velocity
dependent accelerations arise from magnetic forces, and are thus always
perpendicular to $\mathbf{v}$.

As expected from Eqs. (\ref{PrecessGen}) (and by contrast with the
other effects), the precession of a gyroscope carried by $\bar{\mathcal{O}}$,
Eq. (\ref{PrecessTrans}b) takes a form analogous to the precession
of a magnetic dipole, Eq. (\ref{PrecessTrans}a), if we express $\mathbf{S}$
in the local orthonormal tetrad $e^{\hat{i}}$, non rotating relative
to the inertial observer at infinity, such that $S^{i}=(1-M/R)S^{\hat{i}}$:
\begin{equation}
\frac{d\mathbf{S}}{d\bar{t}}=\frac{q}{2m}\frac{Q}{c^{2}R^{3}}\left[\mathbf{S}\times(\mathbf{R}\times\mathbf{w})\right]\ \ (a);\ \ \ \ \ \ \frac{dS^{\hat{i}}}{d\bar{t}}=\frac{2M}{c^{2}R^{3}}\left[(\mathbf{R}\times\mathbf{w})\times\mathbf{S}\right]^{\hat{i}}\ \ (b)\ .\label{PrecessTrans}\end{equation}
If instead of the gyroscope comoving with observer $\bar{\mathcal{O}}$
(with constant velocity $\mathbf{w}$), we had considered a gyroscope
moving in a circular orbit, then an additional term would arise in
analogy with Thomas precession for the magnetic dipole; for a circular
geodesic that term amounts to $-1/4$ of expression (\ref{PrecessTrans}b),
and we would obtain the well known equation for geodetic precession
(e.g. \cite{Gravitation and Inertia}).

\section{Conclusion}

We conclude our paper by discussing some of the implications of our
conclusions in the approaches usually found in literature. In the
framework of linearized theory, e.g. \cite{Ruggiero:2002,Gravitation and Inertia},
Einstein equations are often written in a Maxwell-like form; likewise,
geodesics, precession and gravitational force on a spinning test particle
are cast (in terms of 3-vectors defined in analogy with the electromagnetic
fields $\mathbf{E}$ and $\mathbf{B}$) in a form similar to, respectively,
the Lorentz force on a charged particle, the precession and the force
on a magnetic dipole. 

We have concluded that the actual physical similarities between gravity
and electromagnetism (on which the physical content of such approaches
relies) occur only on very special conditions. For tidal effects,
like the forces on a gyroscopes/dipoles, the analogy manifest in Eqs.
(\ref{FG_Stationary}) holds only when the \emph{test particle} sees
time independent \emph{fields}. In the example of analogous systems
considered in section \ref{Translational-vs.-Rotational}, this means
that the center of mass of the gyroscope/dipole must not move relative
to the central body. In the case of the analogy between the equation
of geodesics and the Lorentz force law (see Fig. \ref{fig1}), as
manifest in equation (\ref{geoGeneral}), it is in the \emph{potentials/metric
perturbations}, as seen by \emph{the observer} (not the test particle!),
that the time independence is required. The latter condition is not
as restrictive as the one of the tidal effects: consider for instance
observers moving in circular orbits around a static mass/charge; such
observers see an unchanging spacetime, and unchanging electromagnetic
potentials, so, for them, the equation of geodesics and Lorentz force
take similar forms (such analogy may actually be cast in an exact
form, see \cite{Natario,Jantzen}). However, those observers see a
time-varying electric field $\mathbf{E}$ (constant in magnitude,
but varying in direction), which, by means of equations (4) and (6),
implies that the tidal tensors are not similar to the gravitational
ones%
\footnote{The electromagnetic field $F^{\alpha\beta}$ is not constant along
the worldline of an observer moving in a circular orbit (radius $R$,
angular velocity $\bm{\Omega}$, velocity $\mathbf{w}=\bm{\Omega}\times\mathbf{R}$)
around a point charge. Its variation endows the magnetic tidal tensor
with an antisymmetric part, and the electric tidal tensor with a time
component: $dF^{0i}/d\tau=Qw^{i}/cR^{3}=-2E^{[i0]}=-\epsilon^{ijk}B_{[jk]}$.
This means that they significantly differ from the GR tidal tensors
seen by an observer in circular motion around a point mass.

Note that both the GR and the EM tidal tensors for these analogous
problems can be obtained from, respectively, Eqs. (\ref{Egij})-(\ref{Hij})
and (\ref{Eij})-(\ref{Bi0}), making therein $\mathbf{R}.\mathbf{w}=0$
(corresponding to circular motion), despite the fact that these expressions
were originally derived for an observer with constant velocity. This
is because, as can be seen from their definitions in Table \ref{analogy},
it is the 4-velocity $U^{\alpha}$ (regardless of the way it varies),
at the given point, that determines the tidal tensors. %
}. 

Finally, as a consequence of this analysis, a distinction, from the
point of view of the analogy with electrodynamics, between effects
related to (stationary) rotational mass currents, and those arising
from translational mass currents, becomes clear: albeit in the literature
both are dubbed {}``gravitomagnetism'', one must note that, while
the former are clearly analogous to magnetism, in the case of the
latter the analogy is not so close.

\vskip 20pt
\leftline{{\normalfont\bfseries Acknowledgments}}
\vskip 4pt
We thank the anonymous referee for very useful comments and suggestions.

\end{document}